\documentclass[reprint,amsmath,amssymb,aps]{revtex4-1}
\usepackage{graphicx}
\usepackage{dcolumn}
\usepackage{bm,romannum}
\usepackage{epstopdf}

\begin{document}

\title{Connection between the mass flow rate and the base and bulk normal stresses in silo discharge}
\author{Ashish Bhateja}
\email{bhateja@iitb.ac.in}
\affiliation{Department of Chemical Engineering, Indian Institute of Technology Bombay, Powai, Mumbai 400076, India}

\date{\today}
\begin{abstract}
The discharge of polydisperse grains in a two-dimensional silo, operating in a continuous-discharge mode, is studied with the help of soft-particle discrete element simulations. We find that the mass flow rate displays similar variation with vertical normal stress at the base and the transition-point in the bulk, signifying that the base normal stress can be considered as a representative of its bulk counterpart. The variation of the base and transition-point normal stresses with fill height follows the Janssen's model, with the former being larger than the latter. The transition-point ($y_t$) is defined as the vertical extent of Region of Orifice Influence (ROOI), which is situated directly above the orifice in its neighbourhood. The transition-point occurs largely at the same location irrespective of the fill height. It shifts, however, upon changing the orifice size, indicating its occurrence to be a localized phenomenon. Finally, the scaling of the vertical velocities at the transition-point and outlet uncovers $y_t$ to be a more relevant length scale than the orifice size $D$. This scaling provides a new insight into the dynamics of the silo discharge in the case where the flow rate varies but the normal stress stays largely invariant to change in the orifice size.
\end{abstract}

\maketitle

\section{Introduction}

The silo is an important device used for handling granular materials in various industrial applications. A thorough understanding of granular flow in a silo is crucial for efficient material processing in many, especially, pharmaceutical, food-processing and agricultural industries. In a draining silo, a granular assembly exhibits three distinct flow regimes, namely, quasi-static zones near the corners, rapid flow adjacent to the outlet and the dense flow region upstream of the rapid flow regime \citep{vidyapati2013}. The coexistence of these three regimes makes the analysis of silo discharge process complex.

Granular materials display interesting features in a silo in both static and dynamic situations. For instance, the vertical normal stress saturates below a certain depth in static granular assemblies -- a phenomenon popularly known as the Janssen effect \citep{sperl2006,landry2004,vanel1999}. In fact, this property remarkably differentiates granular matter from liquids, where pressure increases linearly with the depth \citep{jaeger1996}. Interestingly, the Janssen effect is retained even when the side walls are put into vertical motion \citep{bertho2003}. However, hydrostatic behaviour is observed when the walls are vibrated horizontally \citep{martinez2008}.

Many striking characteristics appear when grains are allowed to fall from an orifice located at the bottom of a silo. In case of flowing grains, several studies report the independence of flow rate on the fill height beyond its critical value \citep{nedderman1982,anand2008,ahn2008,perge2012,staron2012}. A legitimate question, in such a case, is then to ask whether the flow rate has any correlation with the normal stress at the silo base or in the bulk. There is no consensus, however, on this matter \citep{ahn2008,aguirre2010,aguirre2011,perge2012}. The experiments of Aguirre \textit{et al.} \cite{aguirre2010,aguirre2011} and Perge \textit{et al.} \cite{perge2012} report no role of normal stress, measured at the base, in governing the flow rate. On the contrary, the experimental investigation of Ahn \textit{et al.} \cite{ahn2008} presents correlation between the base normal stress and the flow rate. Aguirre \textit{et al.} performed experimental study in a two-dimensional \textit{horizontal silo} utilizing a conveyor belt, moving at a constant velocity, for transporting the disks through an aperture in the silo. Whereas, the rest of the experiments were conducted in a \textit{vertical silo} wherein gravity induces the granular flow. Here, the terms horizontal silo and vertical silo are used in the context of denoting net granular flow normal and parallel to gravity, respectively, in the silo.

All these experiments estimated normal stress on the silo base so as to explore its influence on the mass flow rate as the stress measurements are relatively easy to perform at the base rather than in the bulk. However, it is not clear if the base normal stress is directly related to the flow rate as the silo discharge must be driven by stresses acting in the bulk. Thus, a question that naturally arises is whether the normal stress measured at the base is a representative of the normal stress occurring directly above the orifice in the bulk. In the present study, we aim to explore this aspect by means of simulations utilizing the discrete element method (DEM) \citep{cundall1979}. In DEM, one has complete access to kinematic and dynamic information of all particles at a given time, thereby facilitating in estimating the stresses at the base and in the  bulk. Importantly, we also study the association between the mass flow rate and the base and bulk normal stresses by varying the initial fill height at a fixed orifice size, and vice versa. We finally explore velocity scaling to gain more insight into the granular discharge for the case wherein the orifice size is varied at a constant fill height.
\section{Computational details}
\label{sec:DEM}
\begin{figure}[ht!]
\begin{center}
\includegraphics[scale=0.55]{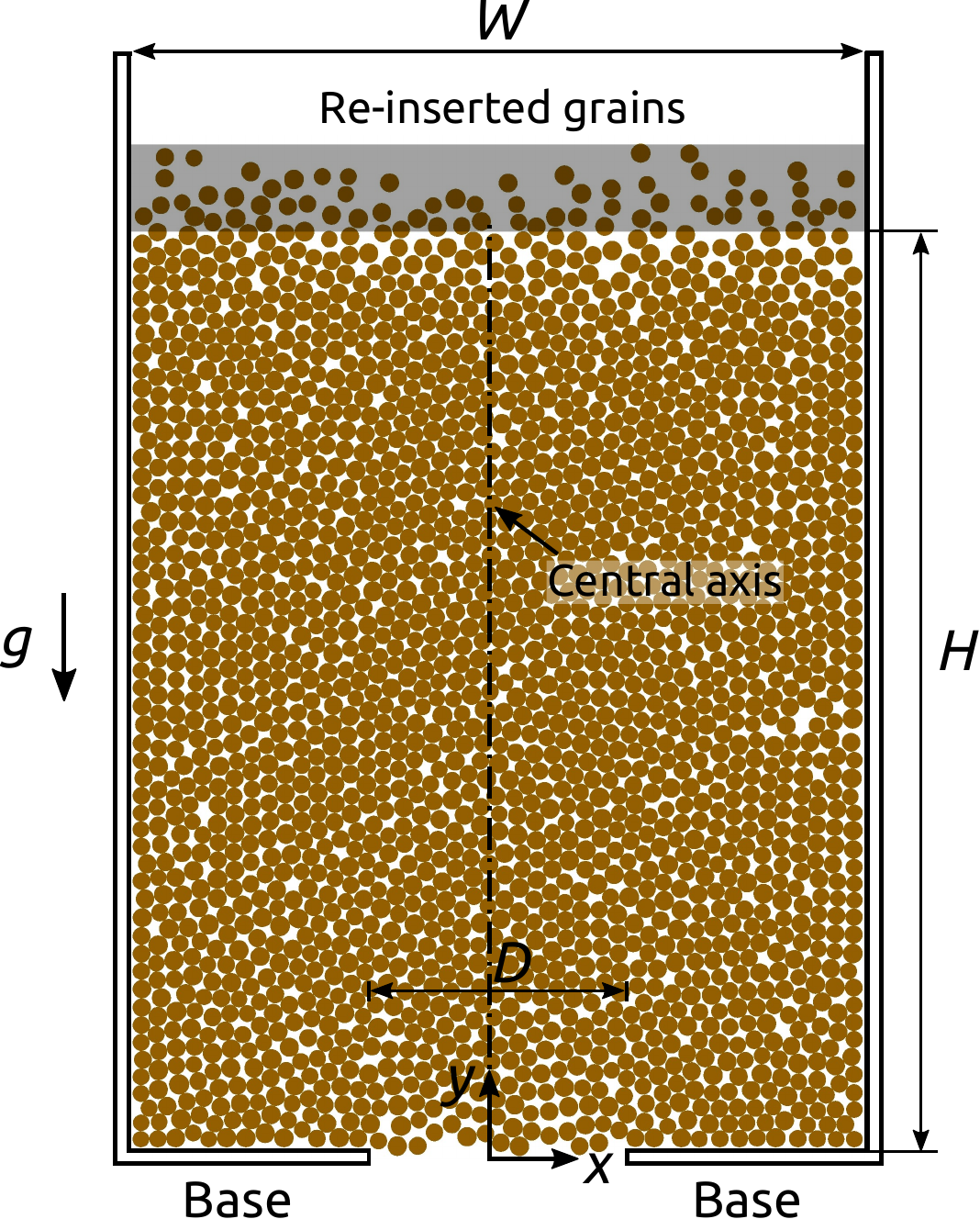}
\caption{A simulation snapshot of granular discharge in a two-dimensional silo for initial fill height $H=50d$; the fill heights up to seven times of it are considered in this study as mentioned in the text. Coordinate axes and direction of gravitational acceleration $g$ are indicated appropriately. Grains are reinserted into the system in the shaded region as displayed on the top.}
\label{fig:ss}
\end{center}
\end{figure}

We consider a two-dimensional vertical silo as shown in Fig. \ref{fig:ss}. The silo operates in the continuous discharge mode in which the exited grains are reinserted into the system at random horizontal locations above the top layer with zero velocity, thus maintaining the fill height largely to its initial value. Moreover, a steady state is achieved for longer duration providing a large statistics for averaging. The grains are modelled as deformable cohesionless disks of mean diameter $d$ with a polydispersity of $\pm 10 \%$ so as to avoid crystallization in the system and its possible consequences on the flow \citep{potapov1996}. The interaction between the grains is modelled through linear spring-dashpot force scheme, considering the grains to be dissipative, with an impact velocity-independent restitution coefficient \citep{cundall1979,zhang1996,bkm2003,shafer1996}. The friction during contact between the grains is incorporated by employing the Coulomb friction criterion \citep{shafer1996}. The same force scheme is employed for interaction between the walls and grains; the details of the force model are provided elsewhere in Bhateja \textit{et al.} \cite{bhateja2016}. The present simulations are carried out with the normal stiffness coefficient $k_n=10^6mg/d$, whereas spring is not considered for tangential direction, i.e., $k_t=0$ and $m$ and $g$ denote the average mass of a grain and gravitational acceleration, respectively. The restitution and friction coefficients between the contacting grains are set at $e_p = 0.9$ and $\mu_p = 0.4$, respectively. The same values are used for wall-grain interactions, i.e., $e_w = 0.9$ and $\mu_w = 0.4$.  The equations of motion are integrated by utilizing the velocity-Verlet algorithm with integration time step $\delta t=10^{-4} \sqrt{d/g}$.

The initial fill height $H$ of the grains is varied between $50d$ and $350d$ in steps of $50d$. Hereafter, the `initial fill height' and `fill height' have the same meaning and will be used interchangeably. The orifice size $D$ is also varied between $9d$ and $14d$ in increments of $1d$. The size $D$ is taken to be larger than $6d$ so as to avoid jamming due to the formation of a stable arch across the orifice \citep{mankoc2007,janda2008,kondic2014}. The silo width $W$ is set at $40d$, ensuring no effect of the side-walls on the flow adjacent to the outlet for the chosen range of $D$, i.e., $W > 2.5 D$ \citep{nedderman1982,vidyapati2013}. The number of particles, $N$, used corresponding to fill height $H$ are given in the format [$H:N$] as follows: [$50d:2150$]; [$100d:4300$]; [$150d:6505$]; [$200d:8700$]; [$250d:10920$]; [$300d:13160$]; and [$350d:15345$].

The data presented in this study are recorded after achieving the steady state and averaged over 50 simulation runs, each beginning with a new initial configuration. The averaging scheme employs coarse-graining technique \citep{goldhirsch2010,weinhart2013,artoni2015}, utilizing a Heaviside step function with coarse-grained width $w$ equal to mean particle diameter $d$. It has been confirmed that the quantities of interest such as velocity, stress and packing fraction do not change upon varying $w$ between $1d-5d$. The stress tensor at a point is computed by summing the stress contribution due to collisions and streaming of grains \citep{campbell1986,goldhirsch2010}. All quantities of interest are made non-dimensional with the mean diameter $d$, density $\rho$ and gravitational acceleration $g$.

\section{Results and discussion}
\label{sec:RD}

We now present the results obtained from discrete element computations. We obtain qualitatively similar results for all fill heights and orifice sizes. Accordingly, our subsequent discussion is based on $H=250d$ and $D=14d$ for the sake of simplicity, unless mentioned explicitly.

As mentioned in the Introduction, we intend to compare normal stress at the base with that of in the bulk. But, the question now is about choosing an appropriate representative point in the bulk to do so. To this end, we resort to the velocity profile, which we discuss next.

\begin{figure*}[ht!]
\begin{center}
\includegraphics[scale=0.42]{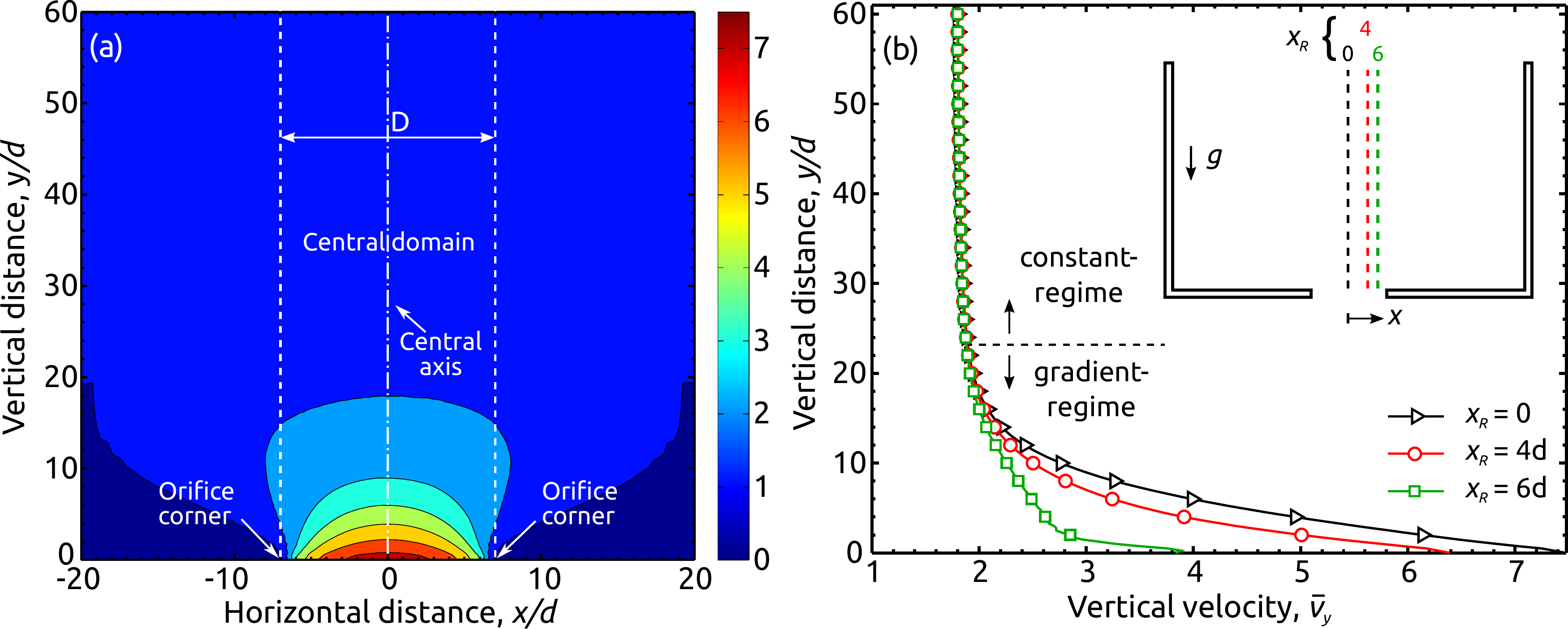}
\caption{(a) Spatial distribution of vertical velocity $\overline{v}_y$. The rectangular region between the dashed lines, labelled as \textit{central domain}, is the area located directly above the orifice. (b) Variation of $\overline{v}_y$ with $y/d$ for different horizontal ($x_R$) locations schematically indicated in the inset. The data are considered for $H=250d$ and $D=14d$, while shown for clarity for a small domain lying between the base ($y=0$) and $y=60d$.}
\label{fig:yvel}
\end{center}
\end{figure*}
%

\subsection{Transition-point}
The spatial distribution of vertical velocity $\overline{v}_y=|v_y|/(gd)^{1/2}$ is examined in Fig.~\ref{fig:yvel}(a), where, $|\cdot|$ denotes the absolute value. For clarity, the distribution is displayed for a small domain up to a height of $60d$ from the base. The velocity distribution is symmetric about the central axis, as expected. The velocities are larger in the region broadly located upstream of the orifice in its vicinity and lowest adjacent to the base and side walls. The velocity is largely constant in the rest of the domain. These observations are consistent with the earlier investigtations, e.g., Staron \textit{et al.} \cite{staron2012, staron2014}.

This can be further illustrated in Fig.~\ref{fig:yvel}(b), which displays the variation of $\overline{v}_y$ with $y/d$ for three horizontal ($x_R$) locations lying within the central domain as indicated in the inset of Fig.~\ref{fig:yvel}(b). The horizontal locations are chosen on right side of the central axis as a similar trend is obtained at equidistant locations on its left side due to geometrical symmetry. Velocity profile exhibits two features that are common to all $x_R$, i.e., $\overline{v}_y$ varies closer to the orifice and becomes constant while climbing up further. We call latter the \textit{constant-regime} and former the \textit{gradient-regime} (see Fig.~\ref{fig:yvel}(b)). The velocity is largely the same in the constant-regime for all $x_R$. These findings indicate that the side walls and base do not affect the kinematics of flow at distant horizontal and vertical locations, respectively, thereby justifying the choice of employing width $W=40d$ for the given $D$ range as mentioned in Sec.~\ref{sec:DEM}. 

\begin{figure*}[ht!]
\begin{center}
\includegraphics[scale=0.33]{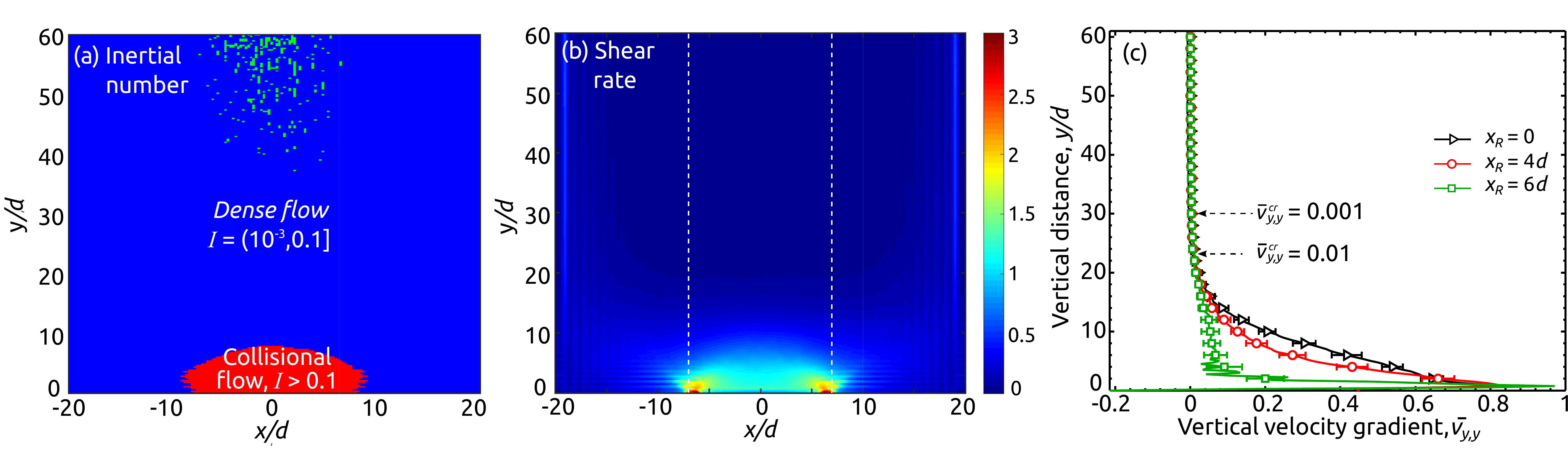}
\caption{(a) Spatial distribution of inertial number $I$. The color scheme as per the range of inertial number is as follows. Red:  $I > 0.1$; Blue: $I=(10^{-3},0.1]$; Green: $I <10^{-3}$. (b) Spatial distribution of shear rate $\overline{\dot{\gamma}}$. The color scale is provided on its right side. (c) Variation of the vertical gradient of the velocity velocity $d\overline{v}_y/d\overline{y}$ for different $x_R$ for the vertical range located between the base and $y=60d$. The data are considered for $H=250d$ and $D=14d$.}
\label{fig:vgrad}
\end{center}
\end{figure*}

In order to further highlight the flow conditions near the orifice, we present in Figs. \ref{fig:vgrad}(a) and \ref{fig:vgrad}(b), respectively, the spatial distributions of inertial number $I=\dot{\gamma}d/\sqrt{P/\rho}$ \cite{midi2004} and shear rate $\overline{\dot{\gamma}} = \dot{\gamma} \, \sqrt{d/g}$. Here, $P$ is the pressure given by the trace of stress tensor $\bm{\sigma}$ and $\dot{\gamma} = (2\bm{G:G})^{1/2}$ with $\bm{G} = 1/2(\nabla{\bm{v}} + (\nabla{\bm{v}})^T)$ being the symmetric part of the traceless velocity gradient tensor $\nabla{\bm{v}}$. In Fig.~\ref{fig:vgrad}(a), the colors (red, blue, and green) are chosen according to the values of the inertial number in the given ranges specified on the corresponding color. Accordingly, two flow regimes, dense and collisional (or rapid), corresponding to the range of inertial number given by \citet{dacruz2005}, are indicated on the plot. A few points (less than $1\%$) shown on the top correspond to $I<10^{-3}$, denoting the quasi-static flow regime \cite{dacruz2005}. Further, it is evident from Fig.~\ref{fig:vgrad}(b) that $\overline{\dot{\gamma}}$ varies significantly in proximity to the outlet indicating the existence of the rapid flow, thereby corroborating well with velocity and inertial number in Figs.~\ref{fig:yvel}(a) and \ref{fig:vgrad}(a), respectively.

We now label the region wherein the effect of the orifice is predominant as \textit{Region of Orifice Influence} (ROOI), spreading largely within the central domain (see Fig.~\ref{fig:yvel}(a)). In fact, the gradient-regime in the central domain is ROOI. The vertical extent of ROOI from the orifice is basically the interface of the constant-regime and gradient-regime. We formally designate the location of this interface from the orifice  as \textit{transition-point} ($y_{t}$), which is considered to be a representative point in the bulk for normal stress comparison. The significance of $y_t$ will become evident in Sec.~\ref{sec:vscale}, where it will be shown to be a relevant length scale in comparison to orifice size $D$. It is important to remark that the term `transition-point' is not used here in the context of phase transition \citep{aharonov1999}. It signifies only the passage of grains from the constant-regime to ROOI. The criterion to locate $y_{t}$ is set according to the vertical gradient of the vertical velocity, i.e., $\overline{v}_{y,y}=d\overline{v}_y/d\overline{y}= \sqrt{d/g}\,(dv_y/dy)$. Figure \ref{fig:vgrad}(c) presents the variation of $\overline{v}_{y,y}$ with $y/d$ for all $x_R$. The earlier observations of constant-regime and gradient-regime in Fig.~\ref{fig:yvel}(b) corroborate well with Fig. \ref{fig:vgrad}(c). Large values of $\overline{v}_{y,y}$ occur adjacent to the orifice, while it approaches close to zero in the constant-regime.

\begin{figure}[ht!]
\begin{center}
\includegraphics[scale=0.42]{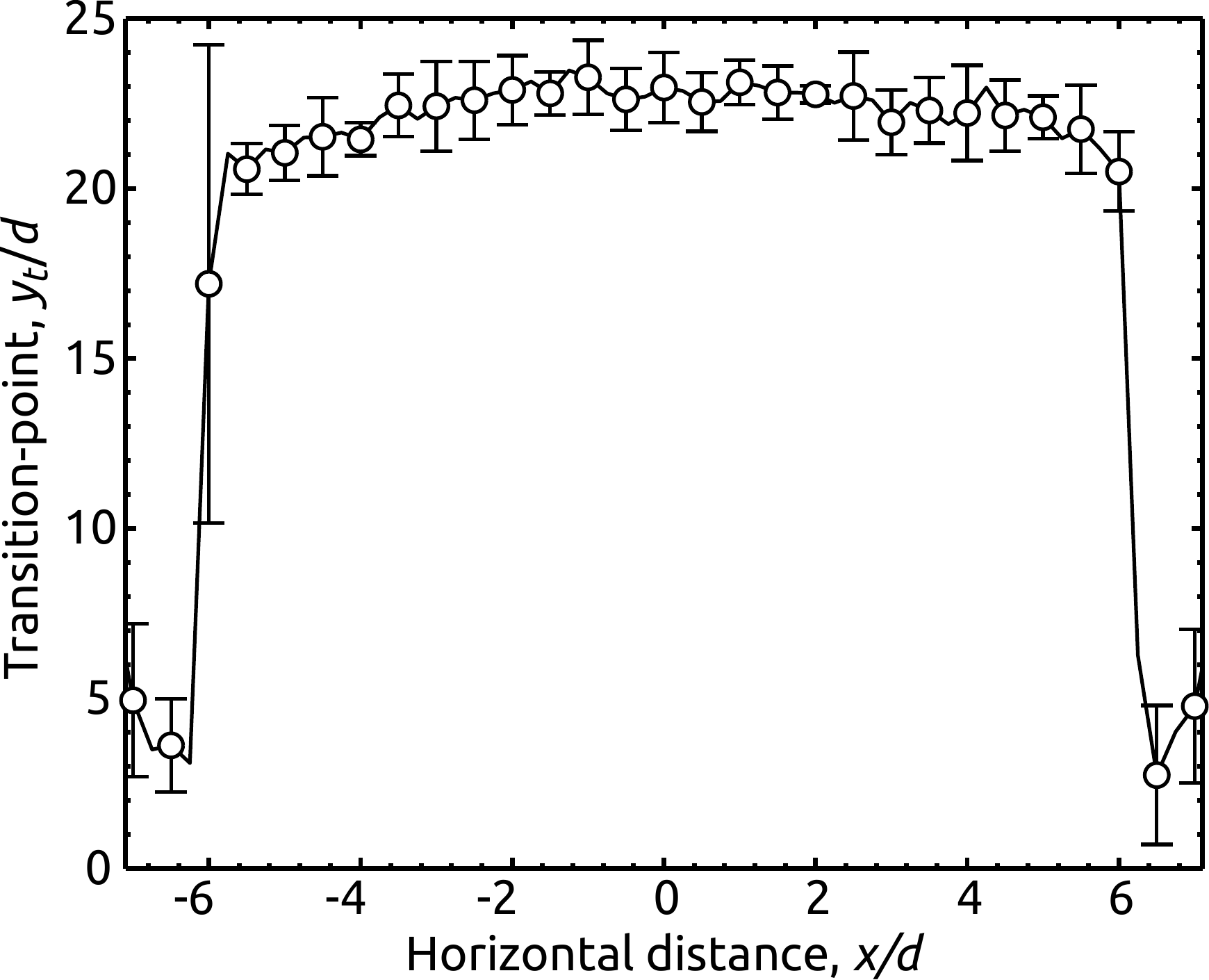}
\caption{Variation of the transition-point $y_{t}/d$ in the central domain with horizontal distance $x/d$, measured from the central axis, for $\overline{v}_{y,y}^{cr}=0.01$. The data are considered for $H=250d$ and $D=14d$.}
\label{fig:tp250}
\end{center}
\end{figure}

We mark the transition-point at a given $x/d$ below which $\overline{v}_{y,y}$ is larger than the critical velocity gradient $\overline{v}_{y,y}^{cr}$. Figure~\ref{fig:tp250} displays variation of $y_{t}/d$ in the central domain with horizontal distance $x/d$, measured from the central axis, for $\overline{v}_{y,y}^{cr}=0.01$. We find that $y_{t}/d$ does not shift appreciably across the central domain, except within $1d$ of its boundary. In other words, the variation in the transition-point is essentially insignificant in the interior of the central domain. It is worth mentioning that another definition of $\overline{v}_{y,y}^{cr}$ gives qualitatively similar results, however, the transition-point shifts upwards for lower $\overline{v}_{y,y}^{cr}$. For instance, on $x_R=0$, $y_{t}\approx 30d$ and $23d$ for $\overline{v}_{y,y}^{cr}=0.001$ and $0.01$, respectively, as indicated in Fig.~\ref{fig:vgrad}(c). But, we note by manual inspection that $\overline{v}_{y,y}^{cr}=0.01$ is a reasonable choice providing us the location closer to the changeover from the constant-regime to ROOI. Whereas, the location indicated for $\overline{v}_{y,y}^{cr}=0.001$ in Fig.~\ref{fig:vgrad}(c) resides in the constant-regime. Therefore, in all what follows, we compute the transition-point for $\overline{v}_{y,y}^{cr}=0.01$.

%

\subsection{Normal stress comparison}
\label{sec:ns}

\begin{figure}[ht!]
\begin{center}
\includegraphics[scale=0.42]{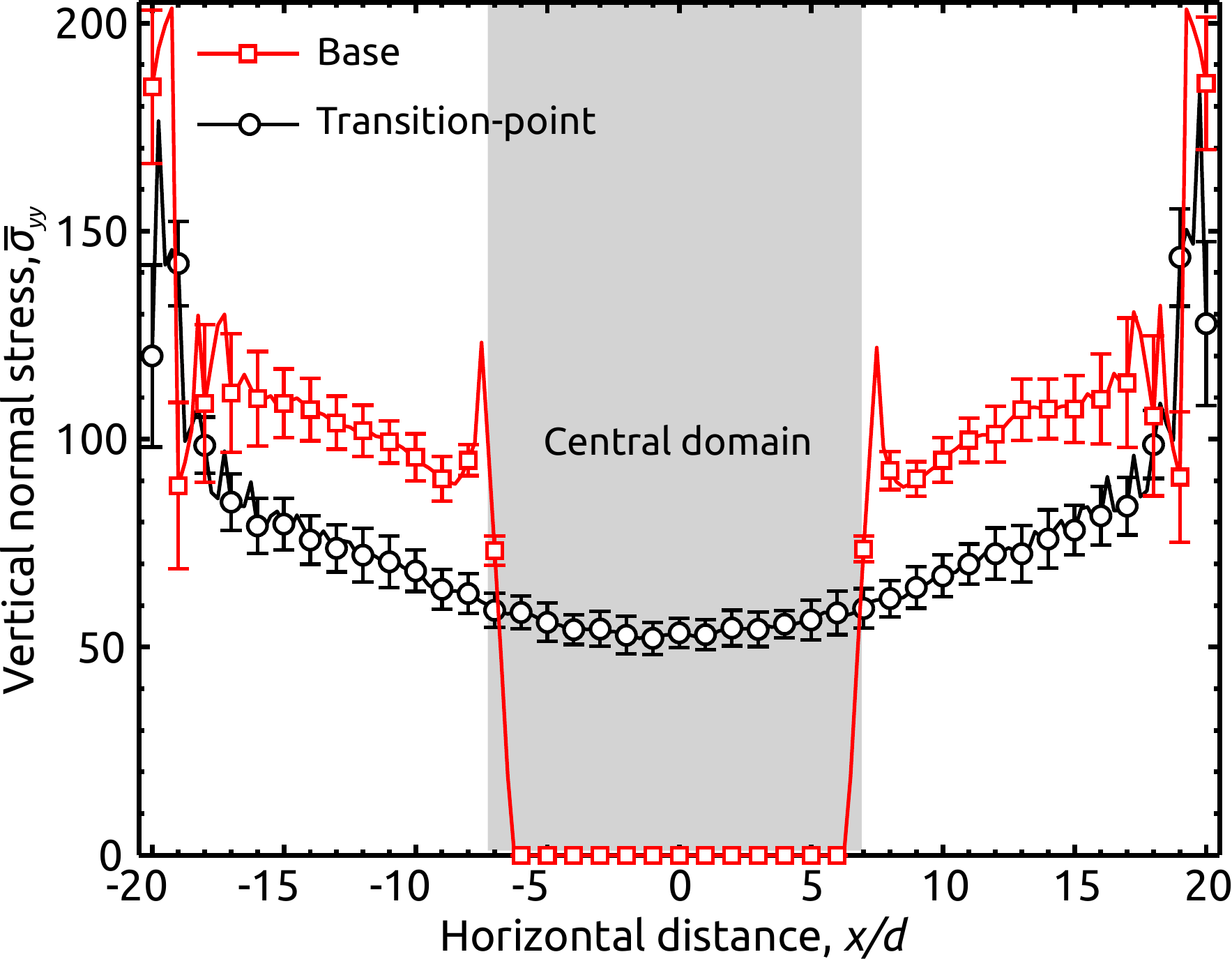}
\caption{Comparison between $\overline{\sigma}_{yy}$ computed at the transition-point $y=23d$ and at the base. The data are given for $H=250d$ and $D=14d$.}
\label{fig:bp}
\end{center}
\end{figure}

We now compare the horizontal profiles of vertical normal stress $\overline{\sigma}_{yy}=\sigma_{yy}/\rho gd$ at the transition-point and base. Henceforth, we use the terms `vertical normal stress' and `normal stress' interchangeably. The transition-point is $y_t^{0}\approx23d$ and the superscript `0' denotes the quantities estimated on the central axis. The normal stress at the base is estimated by computing the force exerted by the particles on it. Figure \ref{fig:bp} compares horizontal profiles of $\overline{\sigma}_{yy}$ measured at $y=23d$ and the base. The vertical normal stress at the transition-point varies weakly across the central domain, which correlates well with the observations of Fig.~\ref{fig:tp250} where transition-point does not change significantly. It shows a slow rise while moving towards the side-walls. Similarly, at the base, $\overline{\sigma}_{yy}$ gently increases  and becomes roughly constant in a small region while approaching the side-walls, barring the sharp peaks and fluctuations occurring adjacent to the orifice corners and the walls, respectively. The occurrence of dip and peak in the base normal stress near the side walls is similar to what is observed in the experiments of Perge \textit{et al.} (2012). Nevertheless, there is a clear difference between $\overline{\sigma}_{yy}$ at the transition-point and the base.

One important notion that appeared in the discussion till now is occurrence of the transition-point denoting the vertical extent of ROOI. The emergence of ROOI seems similar to the picture of \textit{free-fall arch}, enveloping the orifice, wherein the grains fall freely under gravity and $\overline{\sigma}_{yy}$ is assumed to vanish on its boundary \citep{nedderman1982,nmbook}. However, the existence of the free-fall arch is questioned in a recent investigation in view of the absence of zero normal stress in the silo \citep{rubio2015}. The present study also finds that $\overline{\sigma}_{yy}$ does not vanish anywhere, even close to the outlet, thus agreeing with the findings of Rubio-Largo \textit{et al.} \cite{rubio2015}.

\subsection{Transition-point: Initial fill height}
\label{sec:tpH}

\begin{figure}[ht!]
\begin{center}
\includegraphics[scale=0.42]{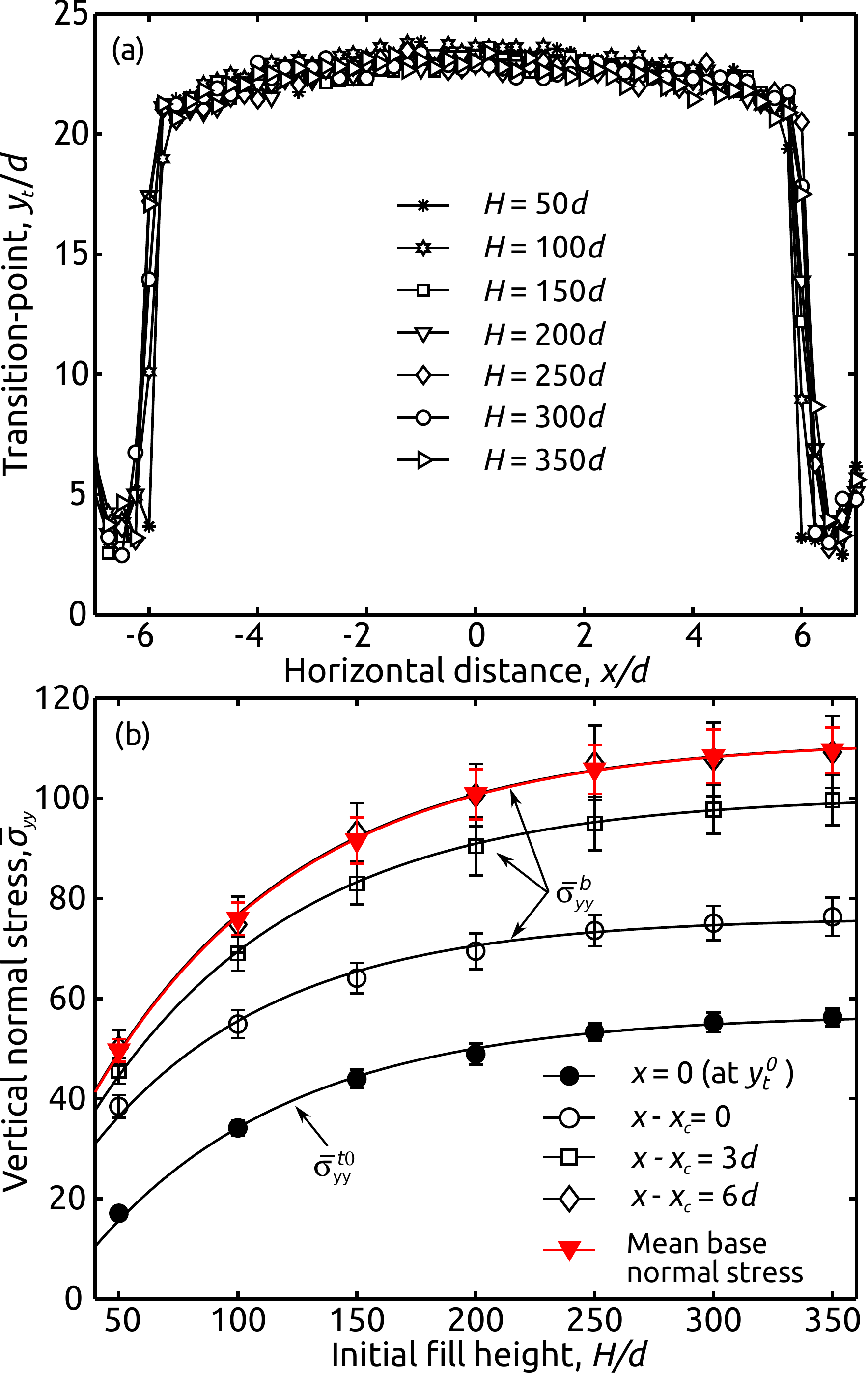}
\caption{(a) Horizontal variation of the transition-point $y_t/d$ for different fill heights in central domain. Error bars are not shown for clarity. (b) Variation of $\overline{\sigma}_{yy}^b$ for three different horizontal ($x$) locations (open symbols) and $\overline{\sigma}_{yy}^{t0}$ (filled circles) with fill height $H/d$. The horizontal locations are given with reference to the orifice corner ($x_c=7d)$. In addition, variation of mean base normal stress with $H/d$ is also shown. Solid lines are fits of Eq.~(\ref{eqn:janssen}). The data are plotted for $D=14d$.}
\label{fig:transH1}
\end{center}
\end{figure}

In this section, we examine how the transition-point varies with the fill height. The results are discussed for $D=14d$ for the sake of simplicity; we obtain similar outcomes for other orifice sizes as well.  Figure~\ref{fig:transH1}(a) displays horizontal variation of the transition-point $y_t/d$ for different fill heights. We note that $y_t/d$ is nearly independent of the fill height. This observation signifies that the fill height does not affect the extent of ROOI, the occurrence of which is a localized phenomenon in the vicinity of the outlet; we show later, however, that the transition-point is indeed sensitive to the orifice dimension. Moreover, $y_t/d$ also does not vary appreciably in the central domain, except near its boundary, as observed earlier in Fig.~\ref{fig:tp250}. Accordingly, hereafter, we compute normal stress at the transition-point on the central axis, $y_t^0$, without loss of generality.
 
We have discussed earlier in Sec.~\ref{sec:ns} that the vertical normal stress at the base $\overline{\sigma}_{yy}^b$ differs with the normal stress at the transition-point $\overline{\sigma}_{yy}^{t0}$, where the superscripts  `b' and `t0' denote that the stress is evaluated at the silo base and the transition-point on the central axis $y_t^0$, respectively. Figure~\ref{fig:transH1}(b) reports variation of $\overline{\sigma}_{yy}^{t0}$, mean base normal stress, and $\overline{\sigma}_{yy}^{b}$ for three different horizontal locations with the fill height. The mean base normal stress is computed by averaging $\overline{\sigma}_{yy}^b$ over the base, excluding the region located within $3d$ of the orifice corners and side walls. As Fig.~\ref{fig:transH1}(b) shows, the normal stress at the base displays similar behaviour as what it does at $y_t^0$, i.e., it rises with increasing the fill height and tend to saturate as the height increases further. However, $\overline{\sigma}_{yy}^b$ is consistently larger than $\overline{\sigma}_{yy}^{t0}$ for all fill heights. There could be several factors contributing to the increased stress at the base \cite{peralta2017,wambaugh2010}. One such factor, for instance, is the initial filling procedure, which is recently reported in the experimental investigation of \citet{peralta2017} employing the batch-discharge mode, i.e., discharge occurs till the silo empties. However, our system differs from their experiments as our silo operates in the continuous-discharge mode, which eliminates the possibility of the influence of initial preparation. On the other hand, the continuous-discharge mode seems similar to the distributed filling protocol of \citet{peralta2017} in light of the pouring  of the exited grains on the top layer, as mentioned in Sec.~\ref{sec:DEM}. However, we obtain saturation of the mass flow rate and vertical normal stress with increasing fill height, contrasting \citet{peralta2017} wherein the apparent mass, measured at the base, does not show saturation with fill height for the distributed filling. Our observation of the saturation of the normal stress and mass flow rate is, nevertheless, similar to what they reported for the concentric filling procedure. Therefore, in light of the above, it appears that the characteristics of both filling protocols, besides the role of force networks observed in the static silo \cite{wambaugh2010}, are likely to be responsible for the increased stress at the base. This requires a separate detailed investigation, which is beyond the scope of the current work.

The variation of the base and transition-point normal stresses with fill height is similar to the Janssen effect observed in static granular assemblies \citep{sperl2006}. Accordingly, the vertical normal stress in a draining silo may be expressed in terms of the fill height and silo width as \citep{sperl2006,peralta2017}
\begin{equation}
\overline{\sigma}_{yy} = \frac{\overline{\gamma} W}{2d\mu_w K} \left(1-e^{-2\mu_w K H/W}\right),
\label{eqn:janssen}
\end{equation}
where $\overline{\gamma}=\gamma/\rho g$ with $\gamma$ being the specific weight and $K$ is the ratio of horizontal to vertical normal stresses. In order to estimate $K$, we consider height to be equal to $H-y_t^0$ and $H$ while fitting the data to Eq.~(\ref{eqn:janssen}) at the transition-point and base, respectively. Solid lines in Fig.~\ref{fig:transH1}(b) are fits of Eq.~(\ref{eqn:janssen}). An excellent fit is obtained for normal stress at the transition-point as well as at the base. Further, it is worth noting that Eq.~(\ref{eqn:janssen}) fits well to the base normal stress at different locations as well as to its average value. We checked that the fitting parameter $K$ does not vary significantly neither at the transition-point in the interior of the central domain, nor at the base excluding the locations situated within $3d$ of the orifice corners and side walls. The mean values of $K$ at the transition-point and the base are $0.62\pm0.02$ and $0.57\pm 0.01$, respectively. The lower $K$ at the base indicates decrease in the horizontal redirection of the vertical stresses.

\begin{figure}[ht!]
\begin{center}
\includegraphics[scale=0.42]{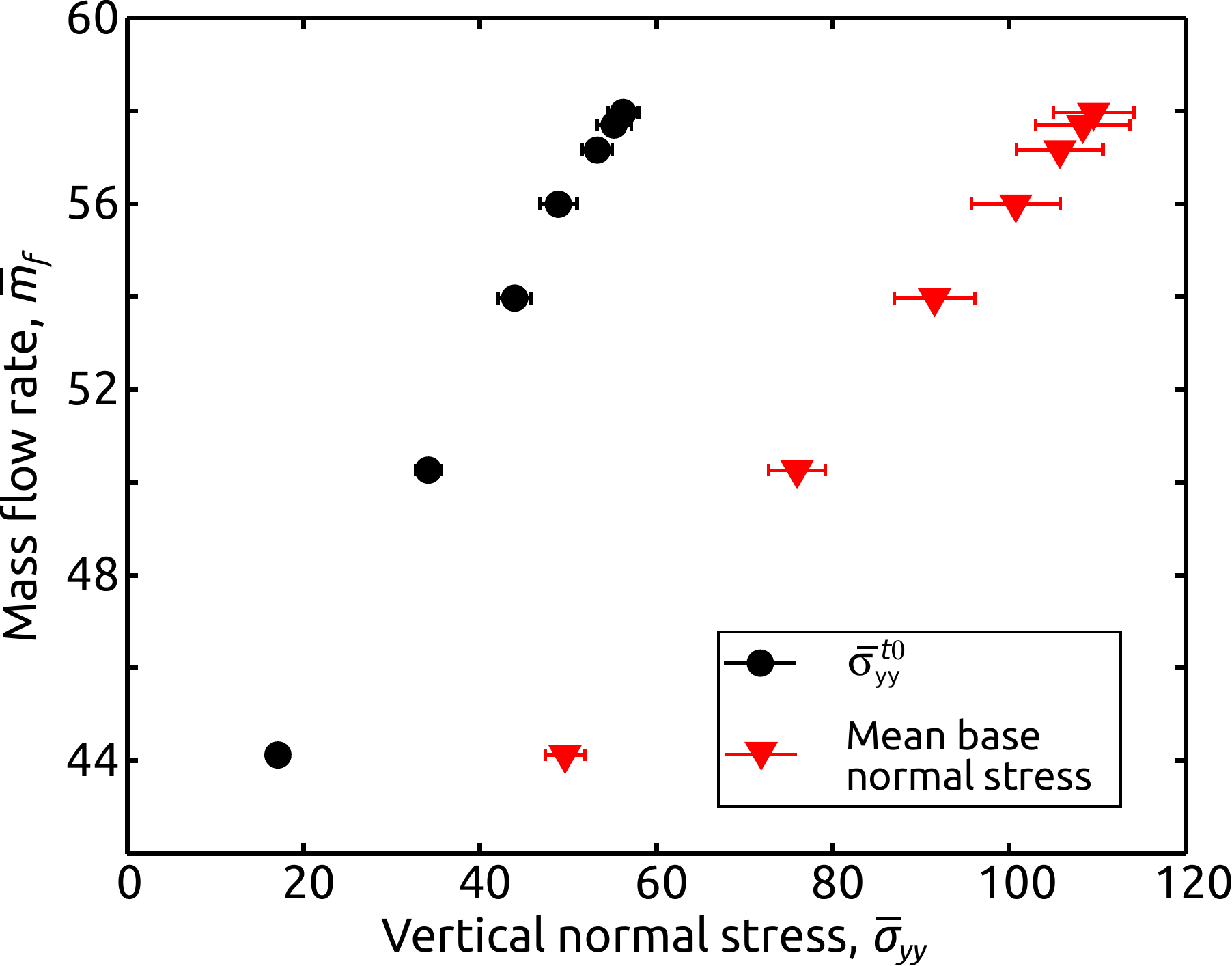}
\caption{Variation of the mass flow rate with the normal stress at the transition-point $y_t^{0}$ and the mean base normal stress. The data are plotted for $D=14d$.}
\label{fig:mfr_st}
\end{center}
\end{figure}

We note in Fig.~\ref{fig:transH1}(b) that the mean base normal stress is almost equal to the stress in the region located a few particle diameter away from the orifice corners and side walls. Therefore, hereafter, we consider the mean base normal stress while comparing it with the transition-point normal stress. The variation of mass flow rate $\overline{m}_f=m_f/(\rho g^{1/2}d^{3/2})$ with $\overline{\sigma}_{yy}$ at $y_{t}^{0}$ and the mean base normal stress is displayed in Fig.~\ref{fig:mfr_st}. The mass flow rate rises as the normal stress increases for both cases, and stays largely unaffected at higher $\overline{\sigma}_{yy}$ corresponding to large fill heights. Note that the mass flow rate also displays similar trend with the base normal stress at different horizontal locations, given the similarity of their profiles with fill height in Fig.~\ref{fig:transH1}(b). The similarity between the curves displaying the variation of the mass flow rate with the base and transition-point normal stresses signifies that the base normal stress, which is typically measured in experiments due to ease, can be utilized as a representative of the bulk normal stress. The existence of this similarity, in fact, justifies measuring the base normal stress in earlier experiments and relating it to the flow rate.

\subsection{Transition-point: Orifice size}
\label{sec:tpD}
\begin{figure}[ht!]
\begin{center}
\includegraphics[scale=0.42]{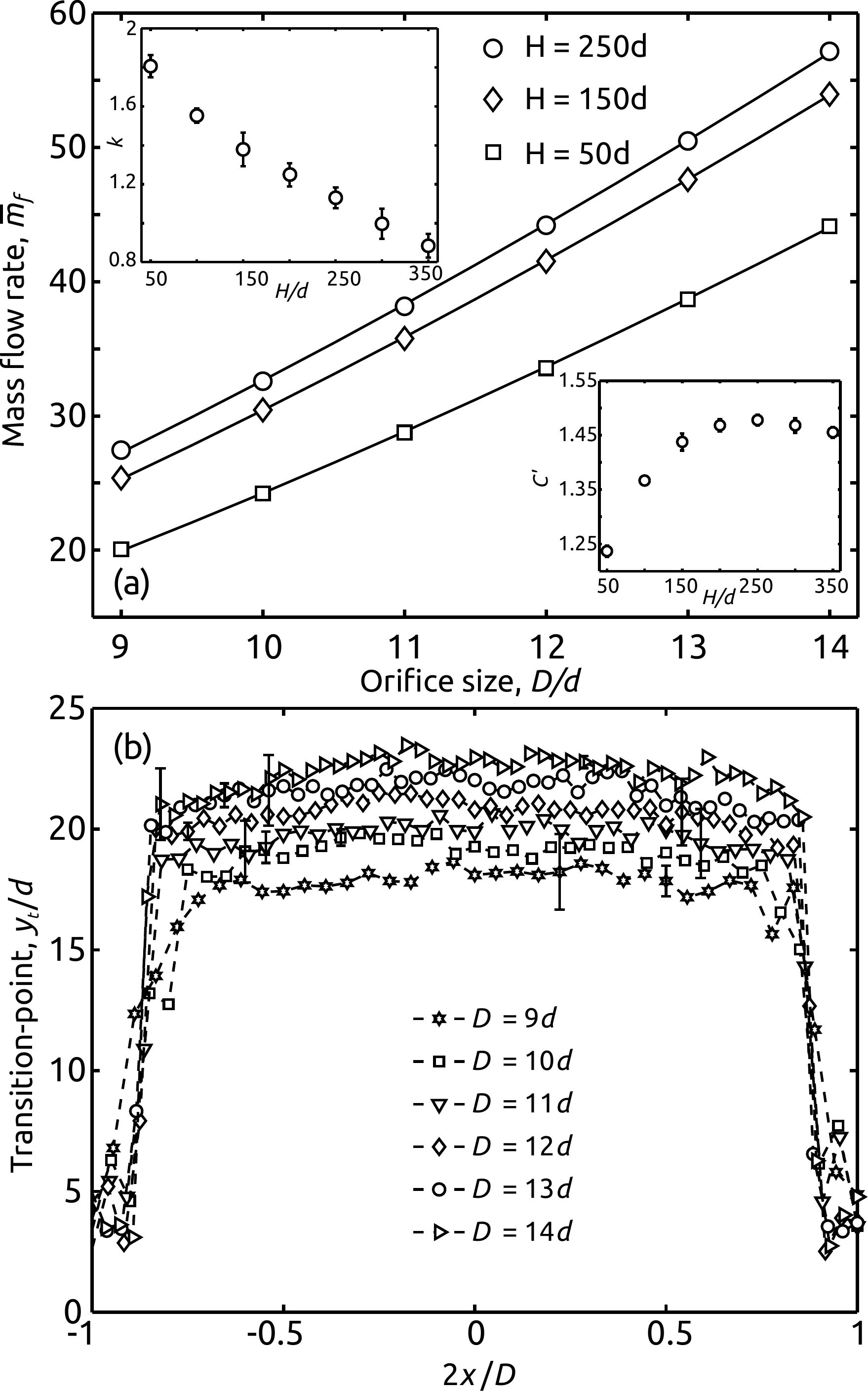}
\caption{(a) Variation of $\overline{m}_f$ with $D/d$ for three different fill heights. Similar behaviour is obtained for other fill heights, which are not shown for simplicity. Solid lines are fits of Eq.~(\ref{eqn:nondim}). \textit{Insets}: Dependence of $k$ and $C'$ on $H/d$. (b) Variation of the transition-point $y_t/d$ for different orifice sizes for $H=250d$. For clarity, the error bars are shown corresponding to their maximum and minimum values for each $D$ in the interior of the central domain.}
\label{fig:mfr_D}
\end{center}
\end{figure}

A commonly used expression relating $m_f$ to $D$ was proposed by Beverloo and co-workers (1961)\nocite{beverloo1961}, which may be written for a two-dimensional silo as \citep{nedderman1982}
\begin{equation}
m_f = C \, \rho_b \, g^{1/2} \, (D-kd)^{3/2},
\label{eqn:dim}
\end{equation}
where $\rho_b =$ $\rho \, \phi_0$ is the initial bulk density with $\phi_0 \approx 0.84$ being the initial packing fraction of the system before discharge, and $C$ and $k$ are fitting parameters. Equation (\ref{eqn:dim}) may be expressed in the dimensionless form as
\begin{equation}
\overline{m}_f = C^{'} \, \phi_0 \, (D/d-k)^{3/2}.
\label{eqn:nondim}
\end{equation}
Figure \ref{fig:mfr_D}(a) shows the variation of $\overline{m}_f$ with $D/d$ for three different fill heights. Expectedly, $\overline{m}_f$ rises with $D/d$ as more material flows out as the opening widens. The flow rate is observed to follow the Beverloo scaling (Eq.~\ref{eqn:nondim}). The fitting parameters are estimated for each fill height, and their variation is shown in the insets of Fig.~\ref{fig:mfr_D}(a). Surprisingly, $k$ decreases monotonically with $H/d$, whereas $C'$ exhibits a non-monotonic variation and becomes roughly constant at large fill heights. 

A weak variation of the fitting parameters with fill height is also noticed by \citet{staron2012} in their computational continuum analysis of two-dimensional silo discharge. In our case, the variation in these parameters is rather significant which may be attributed to the discrete nature of the system under examination. Typically, $k$ is greater than one for discrete systems \citep{beverloo1961,nedderman1982}. However, it presently goes below unity, which is, perhaps, because of the polydisperse grains. Further, the asymptotic value of $C' \approx 1.46$ matches quite well with what is obtained, $1.4$ and $1.48$, by \citet{staron2012} and \citet{staron2014} in their continuum simulations, respectively. However, our $C'$ differs with the value, $1.22$, as reported by \citet{staron2014} for the discrete system. This disparity is because \citet{staron2014} do not employ bulk density in their flow rate expression. We get, in fact, asymptotic value of $C'\approx 1.23$ by not considering the bulk density. In passing, we draw attention to the fact that the Beverloo correlation (Eq.~\ref{eqn:nondim}) describes well the variation of mass flow rate with orifice size without taking normal stress into account. However, as Fig.~\ref{fig:mfr_st} shows, the mass flow rate varies with the normal stress. This finding along with the variation of the fitting parameters with fill height indicate inclusion of fill height as well in the mass flow rate expression, which we believe should, perhaps, come through normal stress. A promise in this direction is evident from a recent work of \citet{madrid2017} wherein they report a differential equation relating the mass flow rate to the pressure (trace of the stress tensor) by considering the energy balance along with the constitutive relation of $\mu$-$I$ rheology \cite{jop2006}.

We now examine how the transition-point behaves when the orifice size changes. The results are discussed here for $H=250d$ as other fill heights exhibit similar outcomes. Figure \ref{fig:mfr_D}(b) plots the horizontal profiles of $y_t/d$ for different orifice sizes in the central domain. Again, the variation in $y_t/d$ is minimal across the central domain, except near its boundary, for all sizes. Importantly, the transition-point rises with increasing $D/d$, demonstrating that the vertical extent of ROOI depends upon the orifice dimension. We also note that the span of ROOI is larger than the size of the orifice as $y_t/d$ is greater than $D/d$ for all cases.

\begin{figure}[ht!]
\begin{center}
\includegraphics[scale=0.42]{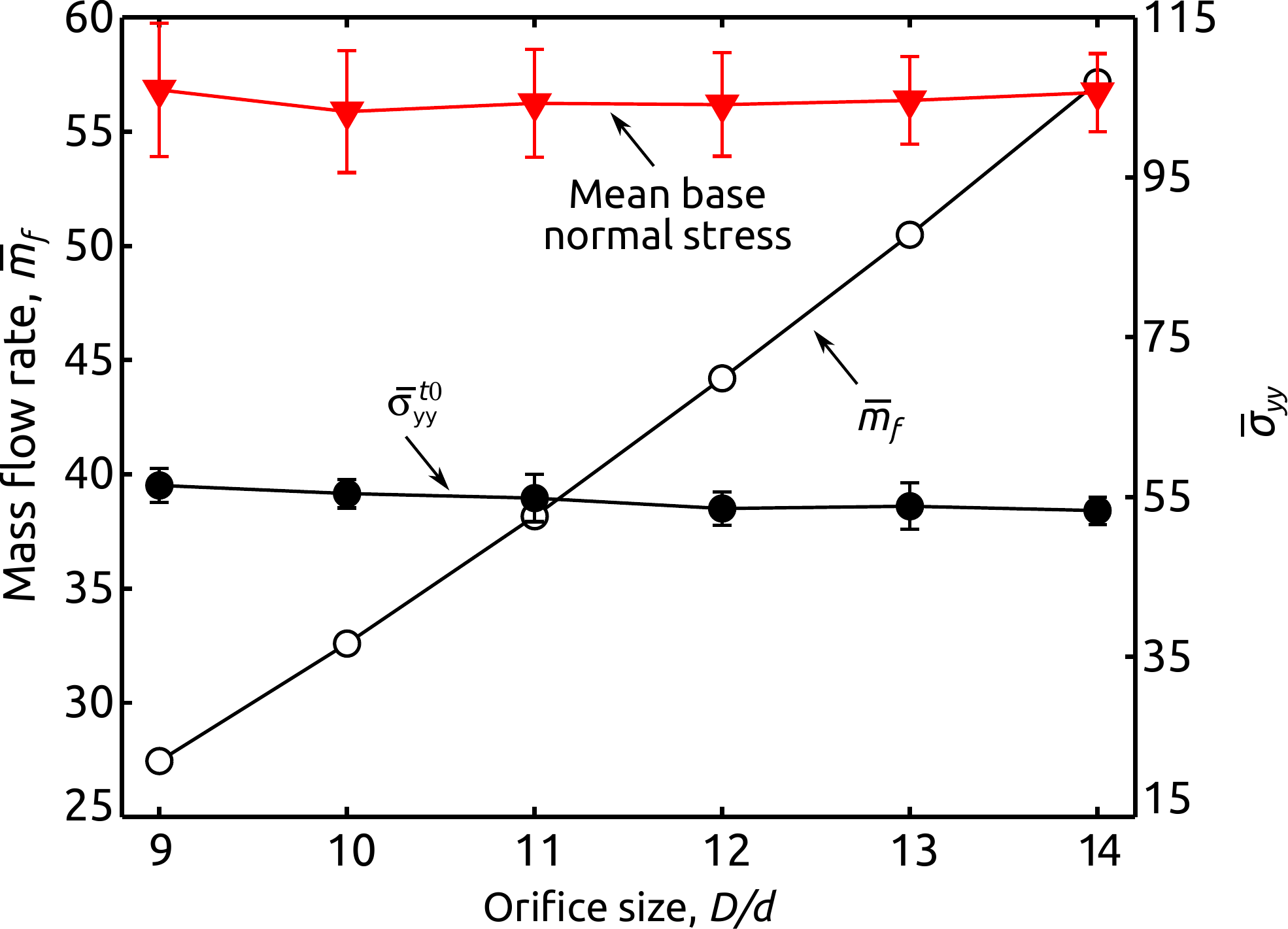}
\caption{Variation of $\overline{m}_{f}$, $\overline{\sigma}_{yy}^{t0}$ and the mean base normal stress with $D/d$.  The error bars in $\overline{m}_f$  are smaller than symbols. Data are given for $H=250d$.}
\label{fig:transD2}
\end{center}
\end{figure}

Figure~\ref{fig:transD2} displays the variation of the transition-point stress $\overline{\sigma}_{yy}^{t0}$, mean base normal stress and mass flow rate with $D/d$. We see that both $\overline{\sigma}_{yy}^{t0}$ and mean base normal stress remain roughly constant as the orifice size changes. Note that the mean base normal stress is largely invariant to change in $D/d$ for other fill heights as well, except for $H=50d$ (not shown for brevity) where it exhibits a slight increase with orifice size. Specifically, the variation is about $10\%$ over a nearly $55\%$ change in the orifice size. The finding of constant normal stress is similar to the observation of the independence of the trace of contact stress tensor on orifice size as reported by Rubio-Largo \textit{et al.} \cite{rubio2015}. Further, as Fig.~\ref{fig:transD2} shows, $\overline{m}_f$ grows with increasing $D/d$. The rise in $\overline{m}_{f}$ with $D/d$, thereby, indicates no relationship between $\overline{m}_f$  with $\overline{\sigma}_{yy}^{t0}$ and $\overline{\sigma}_{yy}^{b}$ (see Fig.~\ref{fig:transD2}), which is in contrast to what is displayed in Fig.~\ref{fig:mfr_st} wherein $\overline{m}_{f}$ relates very well to $\overline{\sigma}_{yy}^{t0}$ and mean base normal stress.

We now consolidate the above findings by noting that, at a fixed $H/d$, the transition-point $y_t/d$ and $\overline{m}_f$ vary with $D/d$ whereas the normal stress remains largely constant. The variation of the transition-point indicates changes occurring at the kinematic level. Obviously, the rise in $\overline{m}_f$ with $D/d$ implies increase in the outlet velocity as well. In light of the rise in outlet velocity along with increase in $y_t/d$, a question which arises naturally is how the outlet velocity relates to the velocity at the transition-point. We next examine velocity scaling to explore this which will provide further insight into the dynamics of silo discharge in the case of varying $D/d$ keeping fill height the same.

%

\subsection{Velocity scaling}
\label{sec:vscale}
\begin{figure*}[ht!]
\begin{center}
\includegraphics[scale=0.35]{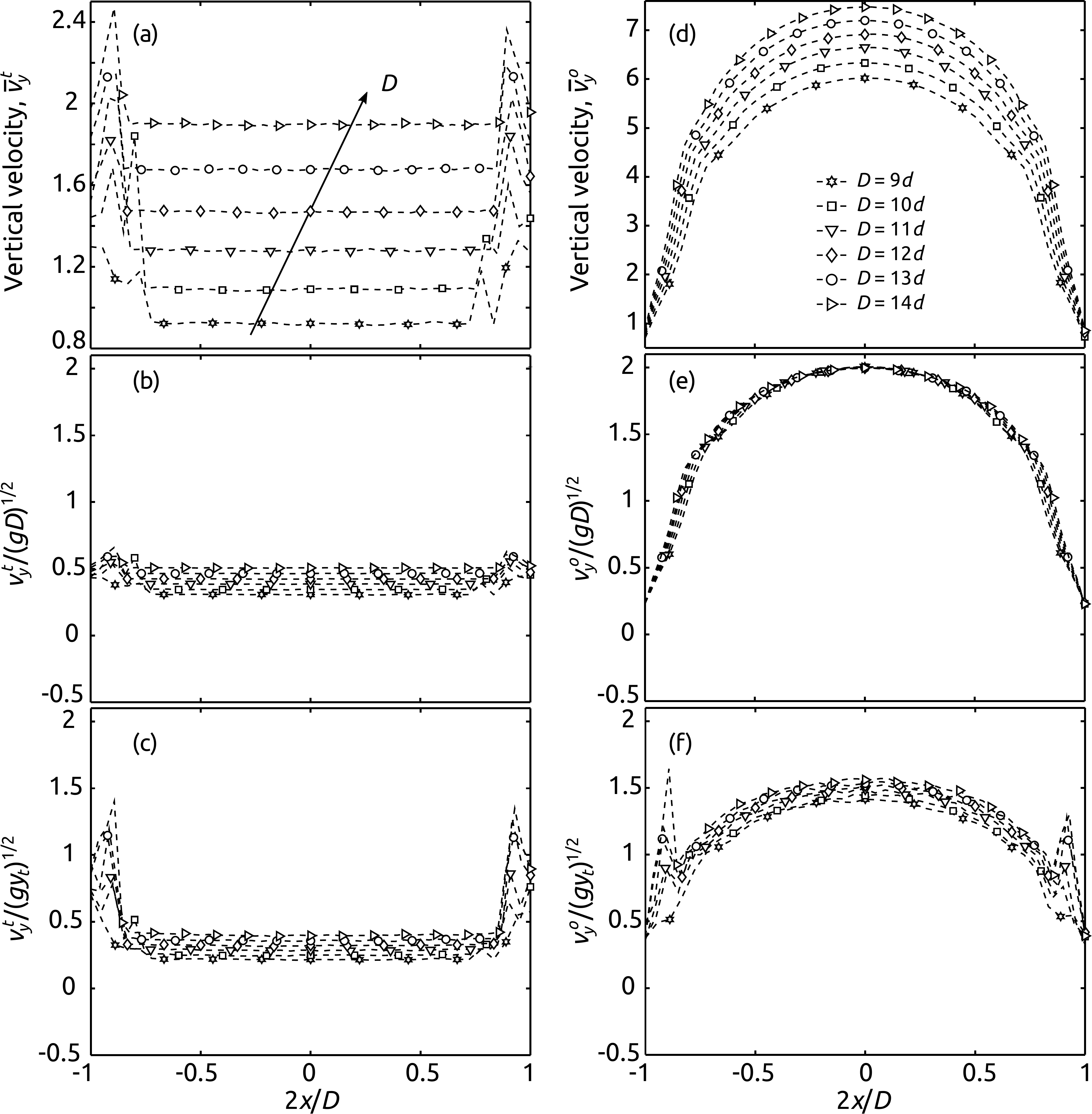}
\caption{(a) Horizontal profiles of $\overline{v}_y^t$ at the transition-point for different orifice sizes. (b) \& (c) Scaling of $v_y^t$ using $D$ and $y_t$ as length scales, respectively. (d) Horizontal profiles of $\overline{v}_y^o$ at the outlet for several $D$. (e) \& (f) Scaling of $v_y^o$ employing $D$ and $y_t$ as length scales, respectively. Legend for all plots is provided in (d). Data are given for $H=250d$.}
\label{fig:vscale}
\end{center}
\end{figure*}

The existence of ROOI in silo discharge provides another length scale $y_t$ besides the orifice size $D$. Therefore, we now explore the scaling of vertical velocity at the transition-point $v_y^t$  with $\sqrt{gy_t}$ and $\sqrt{gD}$ at the fixed fill height $H=250d$. Henceforth, we consider absolute value of the vertical velocity. Figure \ref{fig:vscale}(a) shows horizontal profiles of transition-point velocity $\overline{v}_y^{t}= v_y^t/(gd)^{1/2}$ for different orifice sizes in the central domain. The velocity $\overline{v}_y^{t}$ increases with $D$, however, it remains constant at a given $D$ in the interior of the central domain. We observe in Figs.~\ref{fig:vscale}(b) and (c), respectively, that $v_y^t$ scales with neither $\sqrt{gD}$ nor $\sqrt{gy_t}$. This is not surprising as the occurrence of largely the same stress conditions for all orifice sizes, respectively, at different $y_t$ and the outlet indicates that $v_y^t$ should relate to $y_t$ and the velocity at the outlet $v_y^{o}$. The horizontal profiles of $\overline{v}_y^{o}= v_y^o/(gd)^{1/2}$ for different $D$ are displayed in Fig.~\ref{fig:vscale}(d). Expectedly, $\overline{v}_y^{o}$ increases as $D$ grows. Next, $v_y^{o}$ scales with $\sqrt{gD}$ as shown in Fig.~\ref{fig:vscale}(e), in striking agreement with the earlier reported investigations \citep{janda2012,rubio2015}. However, its scaling with $\sqrt{gy_t}$ presented in Fig.~\ref{fig:vscale}(f) is weak and not prominent as what is noticed with $\sqrt{gD}$ (cf. Fig.~\ref{fig:vscale}(e)). This indicates that the grains do not fall freely under gravity in ROOI, otherwise, $v_y^o$ should have scaled with $\sqrt{gy_t}$ according to purely kinematic arguments.

\begin{figure}[ht!]
\begin{center}
\includegraphics[scale=0.42]{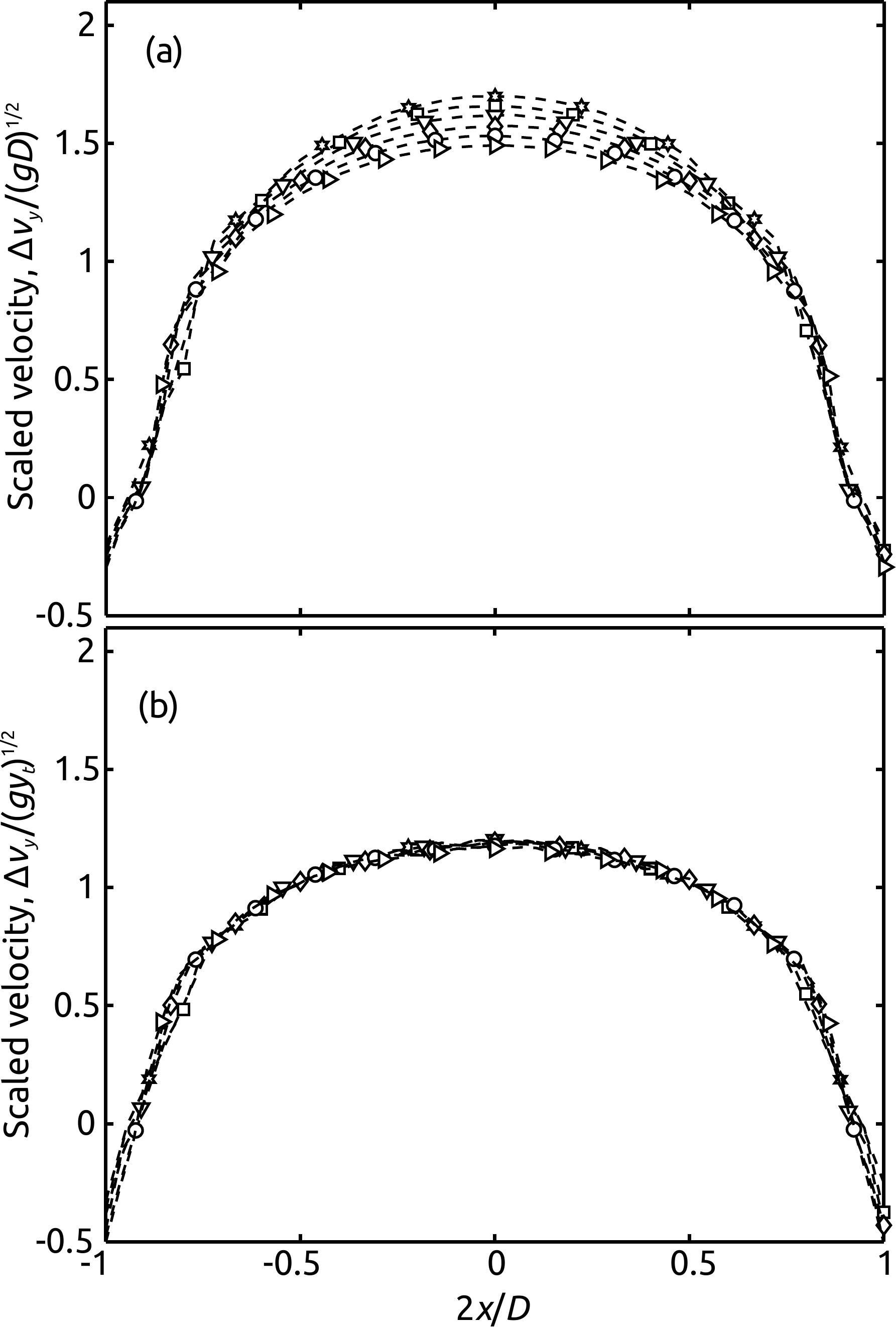}
\caption{Scaling of the relative velocity $\Delta v_y = v_y^{o} - v_y^{t}$ with (a) $D$ and (b) $y_t$ as length scales. Data are shown for $H=250d$.}
\label{fig:vdiff}
\end{center}
\end{figure}

We next examine the scaling of relative velocity $\Delta v_y = v_y^{o} - v_y^{t}$ with both $\sqrt{gD}$ and $\sqrt{gy_t}$. In Fig.~\ref{fig:vdiff}(a), the relative velocity $\Delta v_y$ does not scale with $\sqrt{gD}$ in the middle of the central domain where the difference in $v_y^o$ for successive $D$ is larger (cf. Fig.~\ref{fig:vscale}(d)). However, $\Delta v_y$ scales very well with $\sqrt{gy_t}$ as compared to $\sqrt{gD}$, signifying that $y_t$ is a more relevant length scale for describing the kinematics of granular flow in ROOI. It is important to emphasize that the collapse is, nevertheless, non-trivial in light of the dissimilar shapes of velocity profiles at the transition-point and outlet (cf. Figs.~\ref{fig:vscale}(a) and (d)).

%
\section{Conclusions}

We show in this work that the vertical normal stress at the base $\overline{\sigma}_{yy}^{b}$ and the transition-point in the bulk $\overline{\sigma}_{yy}^{t0}$ vary with fill height $H/d$ in accordance with the Janssen law \citep{sperl2006}, and $\overline{\sigma}_{yy}^{b}$ is higher than $\overline{\sigma}_{yy}^{t0}$. We find that the mass flow rate exhibits similar variation with both base and transition-point normal stresses. This similarity, importantly, suggests that the base normal stress can be employed as a representative of the bulk normal stress, as the former is typically estimated in experiments due to its ease in measurement. The transition-point $y_t/d$ is given by the vertical extent of ROOI from the orifice, which remains largely unaltered by varying $H/d$. It, nevertheless, shifts vertically while changing the orifice size $D/d$, suggesting its occurrence to be a localized phenomenon indifferent to overburden. Moreover, $y_t/d$ remains roughly invariant in the interior of the central domain, the region located directly above the orifice.

In the case of changing $D/d$ at a fixed $H/d$, the vertical normal stress remains largely constant while the mass flow rate and $y_t/d$ vary. The physical insight into the dynamics of silo discharge in this context is gained by examining the scaling of velocities at the transition-point $v_y^t$ and outlet $v_y^o$, considering $y_t$ and $D$ being the length scales. The exit velocity $v_y^o$ scales with $\sqrt{gD}$, in agreement with previous investigations \citep{janda2012,rubio2015}, whereas its scaling with $\sqrt{gy_t}$ is weak. On the other side, $v_y^t$ does not scale with either $\sqrt{gy_t}$ or $\sqrt{gD}$. The relative velocity $\Delta v_y = v_y^{o} - v_y^{t}$ provides a way, which scales very well with $\sqrt{gy_t}$, but not with $\sqrt{gD}$, thereby uncovering the transition-point $y_t$ to be a more relevant length scale than $D$ in describing the kinematics of granular flow in the neighbourhood of the orifice.

\section*{Acknowledgements}
I thank Neeraj Kumbhakarna for providing access to his computing facility for running the simulations presented in this article. I am grateful to Professor Devang Khakhar for insightful discussions and critical reading of the manuscript. Financial support of IIT Bombay is gratefully acknowledged.
\providecommand{\noopsort}[1]{}\providecommand{\singleletter}[1]{#1}%
%
\end{document}